**THz response of nonequilibrium electrons of highly doped graphene on a polar substrate**

S. M. Kukhtaruk

Institute of Semiconductor Physics, NASU, Pr. Nauky 41, Kyiv 03650, Ukraine,
e-mail: kukhtaruk@gmail.com

**Abstract.** High-frequency response of a system with drifting electrons in a highly doped graphene and surface polar optical phonons of a polar substrate is considered. For this interacting system, we obtained a dielectric function, frequencies and decrement/increment of cooperative plasmon-optical phonon oscillations. We found, that the response is significantly depends on the degree of nonequilibrium of the electrons. In particular, interaction between drifting plasmons and surface polar optical phonons leads to an instability of the electron subsystem due to Vavilov-Cherenkov effect. We suggest that the hybrid system – graphene on a polar substrate – is capable to be used in amplifiers or generators of THz electromagnetic radiation.

## 1. Introduction

As is well known [1], graphene shows a number of remarkable physical properties. In particular, a kinetic energy of free carriers as a function of the module of momentum $|\mathbf{p}| = p$ is linear in a wide energy range [2]:

$$E_k = \pm v_F p , \qquad (1)$$

where $v_F \approx 10^8$ cm/s is a Fermi velocity ("+" and "-" in (1) are correspond to electrons and holes respectively). Such symmetry of an energy bands allows considering the graphene as a bipolar semiconductor. To work only with electrons, the electron concentration must be much greater than the concentration of the holes. This can be achieved by doping the graphene. The last can be doped chemically or by applying a positive voltage to a gate, which is generally placed under the substrate. In addition, a such-called electrolytic gate [3] can be used in order to achieve very high electron concentrations.

The substrate influence on the properties of electrons in graphene is studied in [4-10], including the case of nonequilibrium electrons. However, optical properties and co-operative plasma oscillations in highly doped graphene and surface polar optical phonons (SPOP) from a polar substrate are studied only for equilibrium electrons [11-15]. In this paper, the dielectric function of the mentioned above system is calculated in the case of nonequilibrium electrons. Based on this dielectric function, the oscillations of coupled system are studied. The similar studies are reported in [16, 17], but for materials where free electrons have a parabolic energy spectrum.

## 2. Basic equation and model criteria

At first, let us choose a coordinate system and consider an electronic subsystem. Let the $Oz$ axis be perpendicular to the plane of the graphene sheet. Axes $Ox$ and $Oy$ belong to the plane of graphene. The graphene sheet is situated on the polar substrate at $z = 0$. Then, the electronic subsystem will be described using a kinetic equation for the distribution function $F(\mathbf{r}, \mathbf{p}, t)$ of the electrons in highly doped graphene:

$$\frac{\partial F}{\partial t} + v_F \frac{\mathbf{p}}{p} \frac{\partial F}{\partial \mathbf{r}} - e\mathbf{E}\big|_{z=0} \frac{\partial F}{\partial \mathbf{p}} = I\{F\} , \qquad (2)$$

where $\mathbf{r} = (x, y)$ and $\mathbf{p} = (p_x, p_y)$ are the coordinates and momenta respectively, $-e$ is the electron charge ($e > 0$), $t$ is time, and $I\{F\}$ is a collision integral. The total electric field $\mathbf{E}(\mathbf{r}, z, t) = \mathbf{E}_0 + \mathbf{E}_s(\mathbf{r}, z, t)$ consists of the external stationary homogeneous driving electric field $\mathbf{E}_0$ and self-consistent field of the electrons $\mathbf{E}_s(\mathbf{r}, z, t)$. The total field in (2) is calculated in the plane of graphene, i.e. at $z = 0$.

In the case of highly doped graphene, the electron concentration is high and the main mechanism of collision is the electron-electron collision [5, 6, 18, 19]. Then, the stationary spatially uniform solution of the kinetic equation (2) is the shifted Fermi-Dirac distribution function:

$$f_0(\mathbf{p}) = \left( \exp\left( \frac{v_F p - v_0 p_x - \mu}{k_B T} \right) + 1 \right)^{-1} , \qquad (3)$$

where the drift velocity $\mathbf{v}_0 = (v_0, 0)$ is directed along the $Ox$ axis. Here $k_B$ and $T$ are Boltzmann constant and electron temperature respectively. The chemical potential $\mu$ can be determined from the normalization condition:

$$\frac{g_s g_v}{(2\pi\hbar)^2} \int d^2 p f_0 = n_0, \qquad (4)$$

where $g_s = 2$ and $g_v = 2$ are the spin and valley degenerations respectively, $\hbar$ is the reduced Planck constant, and $n_0$ is the equilibrium surface concentration of the electrons.

To describe the electron plasma oscillations we will use the linear approximation, namely: $F(\mathbf{r},\mathbf{p},t) = f_0(\mathbf{p}) + f_1(\mathbf{r},\mathbf{p},t)$, where $f_1(\mathbf{r},\mathbf{p},t)$ is a small perturbation to the function $f_0(\mathbf{p})$. In the linearized kinetic equation for $f_1(\mathbf{r},\mathbf{p},t)$ we will neglect the collision integral:

$$\frac{\partial f_1}{\partial t} + v_F \frac{\mathbf{p}}{p} \frac{\partial f_1}{\partial \mathbf{r}} - e\mathbf{E}_0 \frac{\partial f_1}{\partial \mathbf{p}} = e\mathbf{E}_s\big|_{z=0} \frac{\partial f_0}{\partial \mathbf{p}}. \qquad (5)$$

For convenience, we will use a scalar potential $\varphi(\mathbf{r},z,t)$ instead of the field $\mathbf{E}_s(\mathbf{r},z,t)$, which are connected by the relation $\mathbf{E}_s = -\nabla \varphi$.

Since the equation (5) does not contain the coordinates and time explicitly $f_1(\mathbf{r},\mathbf{p},t)$ and $\varphi(\mathbf{r},z,t)$ can be represented as: $f_1(\mathbf{r},\mathbf{p},t) = f_{\omega,\mathbf{k}}(\mathbf{p})e^{i\mathbf{k}\mathbf{r}-i\omega t+i\cdot 0}$ and $\varphi(\mathbf{r},z,t) = \varphi_{\omega,\mathbf{k}}(z)e^{i\mathbf{k}\mathbf{r}-i\omega t+i\cdot 0}$, where $\omega$ and $\mathbf{k}$ are the frequency and wave vector of the perturbation respectively. The infinitesimal term $i\cdot 0$ arises from the principle of causality and gives Landau rule to bypass the poles in the integrals, which will be given below. The same rule can be obtained by adding the infinitesimal collision integral $f_1/\tau$ at $\tau \to \infty$ to the right side of the kinetic equation (5). Nevertheless, one should keep in mind, that the consideration of the case with a finite value of $\tau$ does not make sense, since such collision integral violates the charge conservation law.

Neglecting of the collision integral is valid if

$$\omega \tau_p >> 1, \qquad (6)$$

where $\tau_p$ is momentum relaxation time of electrons. Using a classical kinetic equation for electrons imposes the following restriction on the value of the wave vector:

$$k << k_F, \qquad (7)$$

where $k_F = \sqrt{\pi n_0}$ is the absolute value of equilibrium Fermi wave vector. If doping of the graphene is carried out by applying a positive voltage to the gate which is located under the substrate then our model is valid if $kd >> 1$, where $d$ is the substrate thickness, i.e. the distance between the gate and graphene sheet.

We also neglect the term $-e\mathbf{E}_0 \frac{\partial f_1}{\partial \mathbf{p}}$ in equation (5). This is valid if changing of the electron momentum under the influence of the field $\mathbf{E}_0$ over the time period of the wave is much smaller than the average electron momentum, and if changing of the electron energy under the influence of this field over the spatial period of the wave is much smaller than the average electron energy (Fermi energy).

As follows from equations (3) and (4), chemical potential decreases with increasing of the drift velocity, so we do not consider the drift velocity of electrons close to $v_F$. If typical speeds of electrons are much smaller than the speed of light, the kinetic equation

(5) should be solved together with the Poisson equation for the Fourier components of the potential $\varphi_{\omega,\mathbf{k}}(z)$, and with necessary boundary conditions:

$$
\begin{cases}
\dfrac{d^2 \varphi_{\omega,\mathbf{k}}}{dz^2} - k^2 \varphi_{\omega,\mathbf{k}} = 0 \\[2mm]
\varphi_{\omega,\mathbf{k}} \mid_{z \to \pm\infty} \to 0 \\[2mm]
\varphi_{\omega,\mathbf{k}} \mid_{z=-\delta} = \varphi_{\omega,\mathbf{k}} \mid_{z=\delta} \\[2mm]
\dfrac{d\varphi_{\omega,\mathbf{k}}}{dz} \mid_{z=\delta} - \varepsilon(\omega)\dfrac{d\varphi_{\omega,\mathbf{k}}}{dz} \mid_{z=-\delta} = \dfrac{4\pi e g_s g_v}{(2\pi\hbar)^2} \int d^2 p\, f_{\omega,\mathbf{k}}(\mathbf{p}),
\end{cases}
\tag{8}
$$

where $\delta \to +0$, and dielectric permittivity $\varepsilon(\omega)$, which included in the latest boundary condition, is a function of frequency (see e.g. book [20]):

$$
\varepsilon(\omega) = \varepsilon_\infty + \frac{(\varepsilon_0 - \varepsilon_\infty)\omega_{TO}^2}{\omega_{TO}^2 - \omega^2 - i\gamma_{TO}\omega},
\tag{9}
$$

where $\varepsilon_0$ and $\varepsilon_\infty$ are the static and high-frequency dielectric constants of the substrate respectively, $\omega_{TO}$ is a frequency of transverse optical phonons. The parameter $\gamma_{TO}$ is describe the phonon damping and $\gamma_{TO} << \omega_{TO}$.

Equations (3)-(5) and (8)-(9) compose a basic system for the problem under consideration.

## 3. Results and discussion

From the basic system of equations, one can obtain a dielectric function for interacting plasmons and substrate phonons

$$
\varepsilon(\Omega,\mathbf{K}) = 1 + \Omega_e \frac{\Omega^2 - 1 + i\Gamma_{TO}\Omega}{\Omega^2 - \Omega_0^2 + i\Gamma_{TO}\Omega} \int \frac{d^2 P}{\left(\Omega - \dfrac{\mathbf{K}\mathbf{P}}{P} + i\cdot 0\right)} \frac{\mathbf{K}}{K} \frac{\partial f_0}{\partial \mathbf{P}},
\tag{10}
$$

where, for convenience, the following dimensionless variables are introduced: $\Omega = \dfrac{\omega}{\omega_{TO}}$,

$K = \dfrac{k v_F}{\omega_{TO}} = \sqrt{K_x^2 + K_y^2}$ , $\Omega_e = \dfrac{4\pi e^2 g_s g_v k_B T}{(2\pi\hbar)^2 v_F \omega_{TO}(1+\varepsilon_\infty)}$, $\Omega_0 = \dfrac{\omega_{SO}}{\omega_{TO}}$, $\Gamma_{TO} = \dfrac{\gamma_{TO}}{\omega_{TO}}$, and $V_0 = \dfrac{v_0}{v_F}$.

The parameter $\omega_{SO} = \sqrt{\dfrac{1+\varepsilon_0}{1+\varepsilon_\infty}}\,\omega_{TO}$ is the frequency of surface optical phonons [13]. Note, that the dielectric function (10) is anisotropic if electron drift is present, so this could lead to a rotation of the plane of the light polarization. Also, if we put $\omega_{SO} = \omega_{TO}$ or $\varepsilon_0 = \varepsilon_\infty$ then we obtain the dielectric function for pure plasma oscillations in highly doped graphene, because, in this case, the width of the residual radiation band is zero.

It is well-known that Landau collisionless damping is absent in highly doped graphene plasma at $k << k_F$. This causes due to the fact that pure plasmon branch does not get to the region where the imaginary part of the dielectric function is non-zero. As was shown for the equilibrium case [12], the interaction of plasmons with SPOP leads to the splitting of the pure plasmon branch into the two branches: $\Omega_+$ and $\Omega_-$. Damping decrement $\Gamma_+$, that corresponds to $\Omega_+$, is equal to $-\Gamma_{TO}/2$, since this branch passes through the region, where imaginary part of integral in (10) is zero. From the other hand, branch $\Omega_-$ passes through the region, where imaginary part of integral in (10) is non-zero. In this region, both $\Gamma_{TO}$ and imaginary part of integral in (10) give contributions to $\Gamma_-$. Thus, for the $\Omega_-$ branch, the nature of damping of the cooperative plasma and SPOP oscillations is the same as the nature of Landau collisionless damping. As will be shown, the similar situation also occur in nonequilibrium case, but starting from some values of drift velocity and $\mathbf{K}$ the damping effect is replacing by growth of oscillations that corresponds to electrical instability of the system.

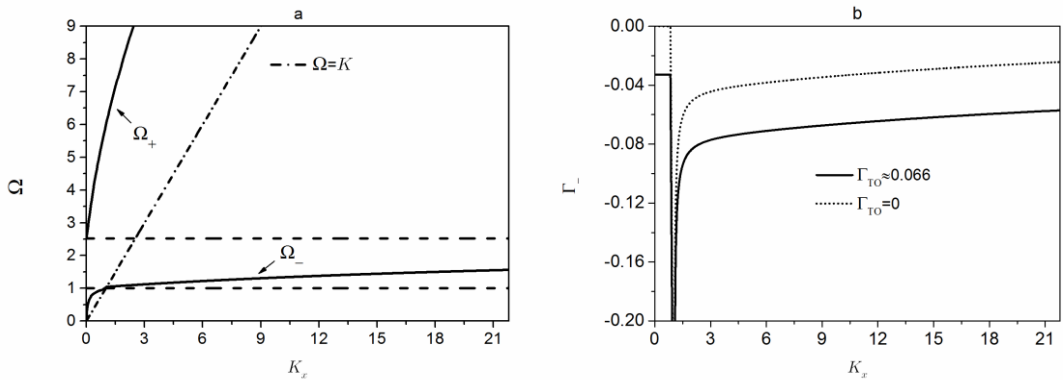

**Fig. 1:** (a) The frequency branches $\Omega_-(K_x)$ and $\Omega_+(K_x)$ at $V_0 = 0$. Dashed lines restrict the residual radiation band. (b) Decrement $\Gamma_-$ as a function of $K_x$ at $V_0 = 0$.

For the numerical calculations we choose TlCl as a material of the substrate. This material [21, 22] has a very wide residual radiation band, since at the helium temperature $\varepsilon_0 \approx 37.2$ and $\varepsilon_\infty \approx 5$. Also, for TlCl $\omega_{TO} \approx 1.15 \cdot 10^{13}$ s$^{-1}$ and $\Gamma_{TO} \approx 0.066$. For the graphene, we take the electron concentration $n_0 = 2 \cdot 10^{12}$ cm$^{-2}$ and electron temperature $T = 400$ K (similar values as in Ref. [6]). For chosen parameters, dimensionless equilibrium Fermi wave vector is $K_F = \dfrac{v_F k_F}{\omega_{TO}} \approx 21.8$. In the case of $V_0 = 0$ the chemical potential is $\mu \approx 153$ meV and $\dfrac{\mu}{k_B T} = 4.4$. If $V_0 = 0.4$ then $\mu \approx 131$ meV and $\dfrac{\mu}{k_B T} = 3.8$.

For simplicity we put $K_y = 0$. The results will be presented for the parameters that satisfy inequalities (6), (7) and other criteria of the model.

As one can see from the Fig. 1a, the frequency branch $\Omega_-$ passes close to $\Omega = 1$ and due to strong interaction this branch never reach the value $\Omega = \Omega_0 \approx 2.5$ (corresponding to $\omega_{SO}$) even at $K_x = K_F$. Since the branch $\Omega_-$ crossing the line $|\Omega| = K$, the damping caused only by phonon damping corresponds to this branch in the region $\Omega > K$ (i.e. $\Gamma_- = -\Gamma_{TO}/2$), and the total damping, caused by phonon damping and Landau damping mechanisms, corresponds to this branch in the region $\Omega < K$. Cutting behavior of $\Gamma_-(K_x)$ can be explained by the fact that the imaginary part of (10) has a point of discontinuity of the second kind at $|\Omega| = K$.

It should also be noted a very rapid growth of $\Omega_+$ branch, and $\Omega_-$ at small $K_x$. Such behavior of the frequency branches leads to high group velocity of the wave in the region of rapid growth. As is well known, group velocity is related to the velocity of energy transfer. Under an applied electric field, much part of the energy is given to almost dispersionless SPOP which group velocity is very small. As a result, the heat is not rejecting from the graphene. In the case of coupled vibrations of graphene plasmons and SPOP, group velocity of waves corresponding to these vibrations is significantly greater. Hence, these waves could provide the heat rejection from the graphene.

In nonequilibrium case (see Fig. 2), both frequency branches differ from those shown in the previous figure not very much. From the other hand, $\Gamma_-$ changes the sign, that is correspond to electrical instability of the system. In absence of the phonon damping the sing of $\Gamma_-$ changes strictly at $\Omega = V_0 K_x$, which corresponds to Vavilov-Cherenkov effect. But, if $\Gamma_{TO}$ is non-zero, the total function $\Gamma_-$ is shifted by the half of the phonon damping parameter $\Gamma_{TO}$. The effect of instability can be used for a generation or amplification of the THz electromagnetic radiation.

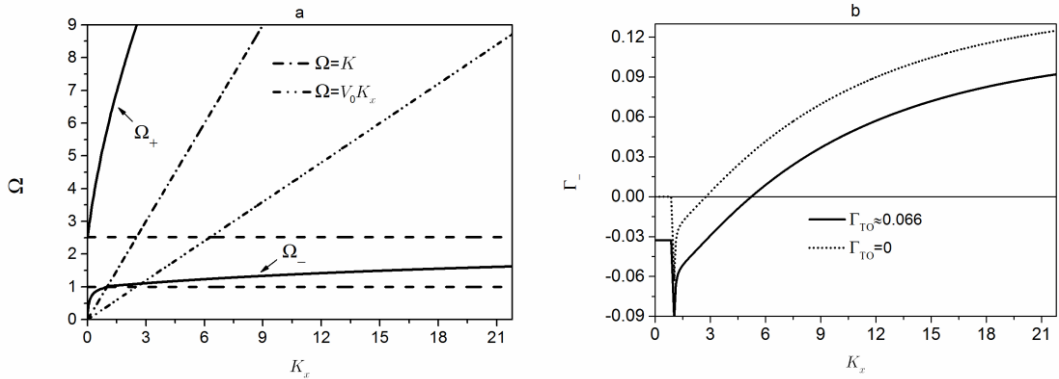

**Fig. 2:** The same as in Fig. 1, but for nonequilibrium electrons with $V_0 = 0.4$.

In the case of presence of the electron drift the electron temperature can be significant higher than the ambient temperature. Our analysis shows, that in the case of significant heating the effect of instability can also occur, if one can create a high enough concentration of electrons.

In conclusion, we have theoretically studied a hybrid system composed by a highly doped graphene and a polar substrate. We have found the THz response of nonequilibrium electrons interacting with surface polar optical pho0ons. The response is significant depends on the electron drift velocity. In particular, if the drift velocity is large, the damping of system oscillations could be replaced by the growth of oscillations. We suggest that the hybrid system under consideration is capable of using for amplification or generation of the THz electromagnetic radiation.

## Acknowledgement

Author is grateful to Prof. V. A. Kochelap for valuable suggestions and constant attention to this work and to Prof. M. V. Strikha for valuable conversation of this paper. The work was partially supported by the State Fundamental Research Fund of Ukraine.